# Enhanced Raman Scattering on Functionalized Graphene Substrates

**Václav Valeš[1], Petr Kovaříček[1], Michaela Fridrichová[1], Xiang Ji[2], Xi Ling[2], Jing Kong[2], Mildred S. Dresselhaus[2], and Martin Kalbáč*[1]**

[1] J Heyrovský Institute of Physical Chemistry, ASCR, v.v.i., Dolejškova 3, 182 23 Praha, Czech Republic

[2] Department of Electrical Engineering and Computer Science, Massachusetts Institute of Technology, Cambridge, Massachusetts 02139, United States

E-mail: martin.kalbac@jh-inst.cas.cz

**Abstract**

The graphene-enhanced Raman scattering of Rhodamine 6G molecules on pristine, fluorinated and 4-nitrophenyl functionalized graphene substrates was studied. The uniformity of the Raman signal enhancement was studied by making large Raman maps. The relative enhancement of the Raman signal is demonstrated to be dependent on the functional groups, which was rationalized by the different doping levels of pristine, fluorinated and 4-nitrophenyl functionalized graphene substrates. The impact of the Fermi energy of graphene and the vibrational energy of the molecules was considered together for the first time in order to explain the enhancement. This approach enables to understand the enhancement of the Raman signal without any assumption about the uniformity of the molecules on the graphene surface and without a necessity to know the enhancement factor. The agreement between the theory and the experimental data was further demonstrated for varying excitation energy.

**1. Introduction**

The concept of graphene-enhanced Raman scattering (GERS) was proposed and experimentally demonstrated in 2010 [1]. It was observed that molecules in very low concentrations exhibit a significant Raman signal when placed on a graphene layer. Furthermore, photoluminescence quenching of molecules on graphene was observed [2], which helps to measure the GERS signal with a reasonable signal/noise ratio for the molecules with strong photoluminescence. GERS is often related to surface-enhanced Raman scattering (SERS) [3], which is often used to obtain Raman scattering signal from an ultralow concentration of measured molecules. SERS is usually explained by the interactions of the light with surface plasmons. This results in high enhancement (up to $10^8$), but the enhancement is not homogeneous. In the case of GERS, the enhancement is lower (10-100), but it is expected to be spatially homogeneous, since graphene can be prepared flat and with a high crystal quality [1]. Since the first observation of GERS, several reports on this effect with various probe molecules [4–6] and even various two-dimensional crystals [7] were carried out.

Explanations of the charge-transfer mechanism in GERS [8] or the effect of dipolar molecular structure [9] have been proposed. The scattering process responsible for GERS has been described by time-dependent perturbation theory by Barros et al. [10]. Within this approach, the following conditions for enhancement have been proposed:

$$(i) \quad \hbar\omega_0 = E_L - E_H, \quad \hbar\omega_0 = E_L - E_H + \hbar\omega_q,$$
$$(ii) \quad E_F = E_H \pm \hbar\omega_q, \quad E_F = E_L \pm \hbar\omega_q,$$
$$(iii) \quad \hbar\omega_0 = E_F - E_H, \quad \hbar\omega_0 = E_F - E_H + \hbar\omega_q,$$
$$(iv) \quad \hbar\omega_0 = E_L - E_F, \quad \hbar\omega_0 = E_L - E_F - \hbar\omega_q,$$

(1)

where $\hbar\omega_0$ is the energy of the excitation radiation, $\hbar\omega_q$ is the energy of the molecular vibration involved, $E_L$ and $E_H$ are the energies of the lowest unoccupied molecular orbital (LUMO) and highest occupied molecular orbital (HOMO) states, respectively, and $E_F$ is the Fermi energy of the graphene sample. The Fermi energy of intrinsic single-layer graphene was reported to be -4.6 eV, and the LUMO and HOMO energies of R6G molecules are -3.4 and -5.7 eV, respectively [1]. Nevertheless, the values of the Fermi level are not of crucial importance for our conclusions as we do not attempt to calculate absolute enhancement. On the other hand, a relative change of the doping is important, because with increasing p-doping of graphene we move closer the HOMO energy of the R6G molecule. It has been shown that the GERS enhancement is the highest in the case when multiple conditions from Equation 1(i-iv) are met [10].

Recently, both SERS and GERS were suggested for applications in chemical and biological sensing devices [11,12]. Graphene thus provides a unique opportunity to realize ultrasensitive sensor devices. However, a general problem of new sensors is their selectivity. Graphene can be chemically functionalized, and thereby sensitivity to specific target molecules can be achieved. Nevertheless, modifying the electronic structure of graphene by functional groups may affect the efficiency of the GERS effect. Consequently, the role of each functional group on graphene in GERS needs to be addressed. Functionalization may lead to charge doping of graphene, which strongly affects the GERS effect. By suitable functionalization, it should be possible to tune maximum enhancement for a selected molecule in a long-term stable manner.

In this work, we used two differently functionalized graphene substrates (fluorinated and 4-nitrophenyl functionalized) and a reference pristine graphene substrate to study the role of functionalization in the GERS effect with Rhodamine 6G (R6G) as probe molecules. With different functionalizations, the Fermi level of the graphene is shifted, thereby affecting the interactions between the molecules and graphene. While 4-nitrophenyl is an electron poor substituent and therefore it should act as a p-dopant of graphene, the doping effect of fluorine atoms depend on the bonding between C and F atoms. In case of partially fluorinated graphene studied here, the doping has been reported not to change significantly [13]. We focused on the enhancement changes depending on the vibrational energy of the R6G molecules and the functionalization of the underlying graphene layer. In contrast to previous studies of GERS effect [5,14,15] we combine here both varying vibrational energy effects and Fermi level energy of graphene in order to the understand the results of GERS measurements. Such approach enables us to validate the theoretical expectations from Equation 1 without knowing an exact amount of R6G molecules on functionalized graphene layer and/or on the Si/SiO$_2$ substrate and even without knowing the enhancement factors. This is an important advantage because of the strong photoluminescence of the R6G molecules on Si/SiO$_2$ substrate, which makes the evaluation of enhancement factor practically impossible.

## 2. Experimental methods

Graphene was prepared by the chemical vapor deposition (CVD) method on copper foil and then transferred onto a silicon wafer with a 300 nm SiO$_2$ layer by a polymer-assisted method (nitrocellulose) described earlier [16]. Grafting with 4-nitrophenyl groups was performed by an adapted literature procedure [17] as follows: 10 mg (42 μmol) of 4-nitrobenzenediazonium tetrafluoroborate was dissolved in 5 ml of deionized water (> 18 MΩ cm$^{-1}$), and the graphene on the silicon wafer was immersed for 30 minutes at room temperature. Then, the sample was removed, washed in 2×5 ml of deionized water and 2×5 ml of methanol (spectroscopy grade), and dried in a stream of argon.

Fluorination was performed using a published protocol [18]. In brief, the graphene on the silicon wafer was fluorinated by XeF$_2$ in the gas phase.

The graphene samples were further soaked in a $10^{-7}$ mol/l aqueous solution of R6G molecules for 2 minutes and subsequently immersed in pure water for 30 minutes to remove possible aggregates of the R6G molecules. (It has been shown that for concentrations below $8 \cdot 10^{-7}$ mol/l a submonolayer coverage of the R6G molecules is formed on a graphene substrate [1] and R6G forms monomers [19].) Atomic force microscopy (AFM) images were taken with the Bruker Dimension Icon using a Scanasyst-air silicon nitride probe. The measurements were performed in PeakForce tapping mode with a peak force set point of approximately 500 pN and a resolution of 512 lines. The images were further processed using Gwyddion software [20].

The resulting Raman maps (20 × 20 points, 2 μm step) were measured with a WITec alpha300 R spectrometer equipped with a piezo stage with a 532 nm excitation wavelength. The laser power was kept below 1 mW to avoid any heating effects. The laser was focused on the sample with a 100× objective to a spot with a diameter of approximately 500 nm. All the Raman peaks were fitted using pseudo-Voigt functions. The pseudo-Voigt lineshape was used because the measured Raman signal is averaged over the laser spot.

The Raman single-spectrum measurements for the wavelengths of 458, 488, 514, 532, 568 and 633 nm (2.71, 2.54, 2.41, 2.33, 2.18 and 1.96 eV) were measured by a Labram HR spectrometer (Horiba Jobin Yvon) with a 100× objective. The measured spot diameter was approximately 1 μm.

## 3. AFM and Raman basic characterization

The samples were characterized by AFM to address the coverage of the graphene layers and the bare Si/SiO$_2$ substrate by the R6G molecules. We measured the same area in graphene before and after deposition of the R6G molecules and we did not observe any signatures of aggregates. The height of the step was found to be around 1 nm and no significant change in the graphene height before and after the deposition of the R6G molecules was found (Figure 1). Therefore, the formation of thick multilayers or aggregates of the R6G molecules on the graphene can be excluded. This observation is in agreement with our previous results from the measurement of the photoluminescence and Raman signal from R6G molecules on graphene [21]. For these reasons, we assume that both graphene and the Si/SiO$_2$ substrate are covered with the R6G molecules.

Typical Raman spectra of the R6G molecules on all the investigated functionalized graphene layers are shown in Figure 2. For further analysis, median values with dispersion described by the first and third quartile obtained from Raman spatial maps were considered. The Raman bands of the R6G molecules are at approximately 610, 780, 1180, 1310, 1350, 1505, 1575, 1590, and 1650 cm$^{-1}$ [2]. The most prominent Raman modes in graphene are the D, G and G´ bands. The D mode is associated with the presence of defects, and therefore it is more pronounced for the functionalized samples. The G mode originates from the doubly degenerate phonon mode in the center of the Brillouin zone, and the symmetry allowed G' mode originates from a second-order process involving two transverse optical phonons [22]. The Raman bands of the Si substrate at 521 cm$^{-1}$ and one centered at 960 cm$^{-1}$ are visible in the spectra as well.

Overall, it can be observed that the Raman signal of the R6G molecules is enhanced on all of the functionalized graphene samples. In our previous work [21] we have shown that the photoluminescence of the R6G molecules is almost completely quenched on both pristine and functionalized graphene substrates. Therefore, we believe that the differences in the observed enhancement of the Raman bands of the R6G molecules can be fully attributed to the GERS effect. Absolutely highest enhancement was measured for 4-nitrophenyl functionalized graphene. However, there are significant differences in the R6G Raman band intensities between differently functionalized graphene samples. This effect might be influenced by modified adsorption ability of the R6G molecules to differently functionalized graphene layers. Therefore, further in the text we focus on variations in the enhancement of different Raman bands of R6G molecules on individual samples. By comparing these experimental results with Equation 1, we are able to draw qualitative conclusions without any assumption on the amount of R6G molecules within the probed area.

By functionalizing graphene we may induce sp$^3$ defects into its structure, which might affect the GERS capability. However, since we observed the highest enhancement for 4-nitrophenyl functionalized graphene sample, we assume that possible decrease of enhancement caused by high ratio of sp$^3$ is negligible compared to effects described in Equation 1. The highest ratio of the amplitude of the D and the G band was observed for the fluorinated graphene and it reached value of around 3. By comparing this result with studies on quantifying defects [23,24], we concluded that the sp$^3$/sp$^2$ ratio is smaller than 10$^{-3}$.

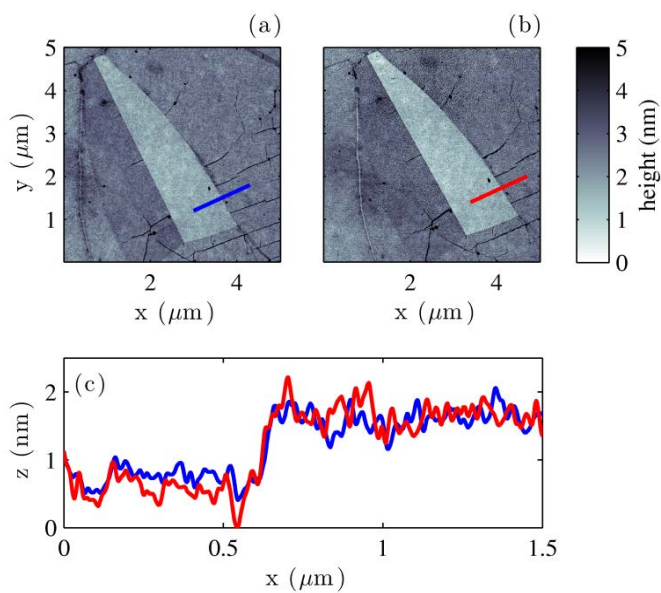

**Figure 1**. AFM maps of an edge of pristine graphene on a Si/SiO$_2$ substrate before (a) and after (b) the deposition of the R6G molecules taken at the same area. Sections across graphene edge without (blue line) and with (red line) R6G molecules are depicted in (c).

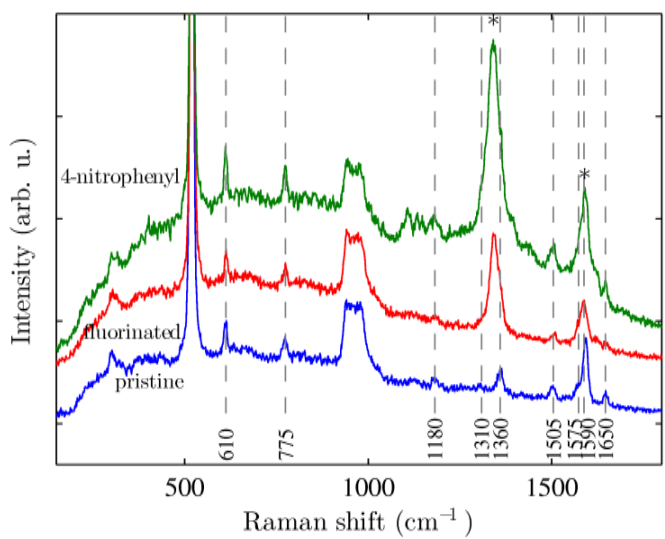

**Figure 2**. Graphene-enhanced Raman scattering of the R6G deposited on differently functionalized graphene layers. The gray dashed vertical lines indicate the positions of the R6G Raman bands. The spectra are vertically shifted for clarity. The measurements were recorded with the excitation wavelength of 532 nm, laser power of 1 mW and the integration time of 5 s.

To obtain statistically relevant data, we measured large spatial Raman maps of the R6G signals on the graphene and Si/SiO$_2$ substrates. The signal from the R6G molecules on the bare Si/SiO$_2$ substrate does not reveal any Raman bands. This is caused by both the absence of GERS enhancement and the strong photoluminescence of R6G molecules. Raman maps of the intensities of the G mode band of graphene and the R6G band at approximately 612 cm$^{-1}$, both related to the intensity of the Si band, are shown in Figure 3. From the maps of the intensity distributions, it can be observed that the variations of the intensity of the R6G Raman band are comparable to the variations of the intensity of the graphene G band. The intensity of the R6G Raman band at 612 cm$^{-1}$ varies on a scale of 40 µm less then by a factor of 2 which indicates a homogenous enhancement of the R6G molecules without any formation of hotspots.

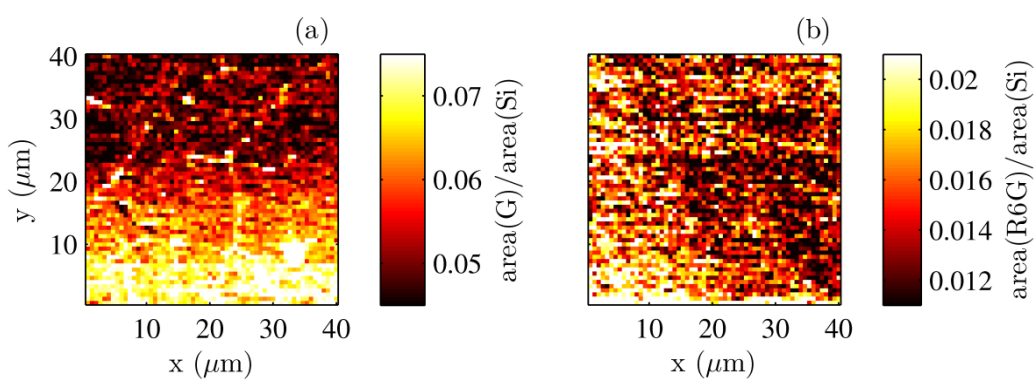

**Figure 3**. Raman maps of the relative areas of the graphene G band (a) and the R6G band at 612 cm$^{-1}$ relative to the area of the Si band at approximately 521 cm$^{-1}$ measured for pristine graphene. The whole measured area is covered by graphene.

## 4. Rationalization of the GERS effect

For a further analysis of the enhancement of the individual Raman modes of the R6G molecules, we focused in detail on the peaks located at approximately 610, 780, 1180, 1505 and 1650 cm$^{-1}$ (that correspond to the vibrational energies of ca. 75, 95, 145, 185 and 205 meV), since the remaining R6G peaks overlap with the D or G bands of graphene. The intensities of the bands at 610, 780, 1180, 1505 and 1650 cm$^{-1}$ are plotted in Figure 4(a) for different functionalizations of graphene. As we did not obtain the Raman signal of R6G on the bare Si/SiO$_2$ substrate, the enhancement factors for the individual bands cannot be calculated. It is, however, possible to discuss the relative enhancement for differently functionalized graphene samples. We chose fluorinated graphene as the reference sample, since it shows the lowest enhancement of the R6G bands (Figure 4(b)). Relative intensities (between differently functionalizes graphene samples) of individual Raman bands of the R6G molecules bring important information since the GERS enhancement is affected besides other things both by the vibrational energy of molecules and the Fermi energy of graphene (Equation 1, Figure 5). Indeed, it can be observed that the relative intensities are highest for the 4-nitrophenyl functionalized graphene and that the relative intensities are increasing with the increasing vibrational energy. The enhancement of the R6G Raman signal of pristine graphene is following the same trend, i.e., the enhancement is higher for higher vibrational energies, but the overall enhancement is smaller compared to that of 4-nitrophenyl functionalized graphene. In Figure 4(c), we related the relative intensities of the analyzed R6G bands to the band with the lowest vibrational energy. From this figure, it is clear that for the highest vibrational energies, the change in the enhancement is higher for pristine graphene than for the 4-nitrophenyl functionalized sample. This observation can be rationalized by the rules of Equation 1 as follow: a) From previous studies of GERS on R6G, it is known that a green 532 nm laser fits the difference between the HOMO and LUMO states of the R6G molecule [1]., b) Different R6G Raman bands correspond to different vibrational energies., c) The Fermi level energy of graphene is further modified by electronic doping. CVD graphene transferred onto a Si/SiO$_2$ substrate under ambient conditions is usually p-doped due to the presence of water between the graphene layer and the substrate [25]. Functionalization may lead to additional doping, which is induced by an interaction between graphene and the functional groups. The doping level was calculated from the correlation analysis of the positions of the G and the G' graphene Raman bands [26,27] measured in the maps. The highest doping was found for the 4-nitrophenyl functionalized graphene ($13.5 \times 10^2$ cm$^{-2}$, 0.36 eV), while the lowest doping was found for the fluorinated graphene ($8.0 \times 10^2$ cm$^{-2}$, 0.28 eV). The pristine graphene showed a doping level of $11.5 \times 10^2$ cm$^{-2}$ (0.33 eV). These values are related to the absolute Fermi energy of graphene of -4.26 eV [28].

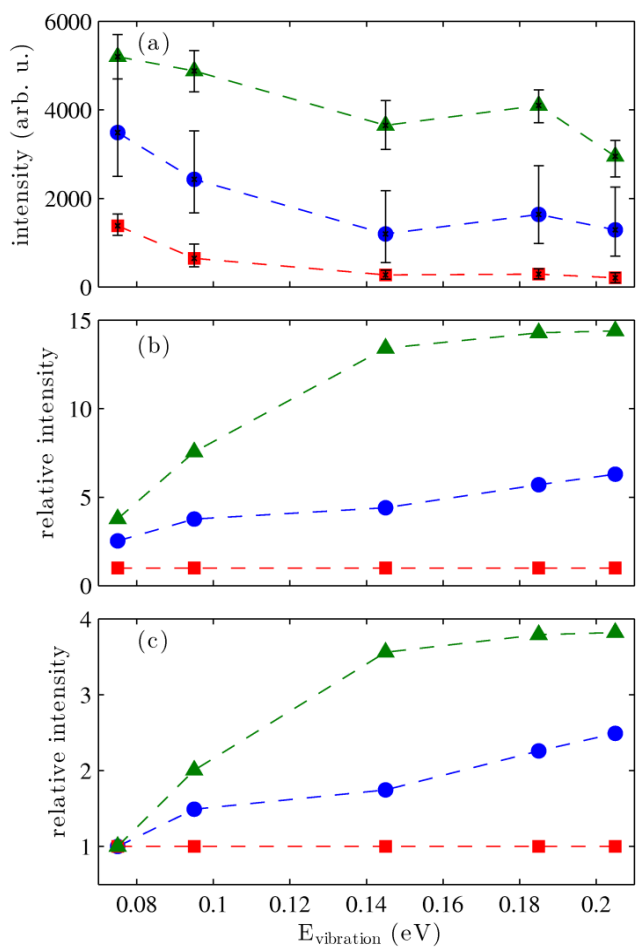

Figure 4. Intensities of different R6G Raman bands for pristine (blue circles), fluorinated (red squares), and 4-nitrophenyl functionalized (green triangles) graphene. Panel (a) shows the absolute intensities of the individual bands. The error bars represent the first and the third quartiles of each datasets from the measured maps. The intensities normalized by the corresponding intensities of R6G on fluorinated graphene are displayed in panel (b). Panel (c) shows the intensities normalized by the intensities of R6G on fluorinated graphene and the intensity of the R6G Raman band with the lowest energy (610 cm$^{-1}$).

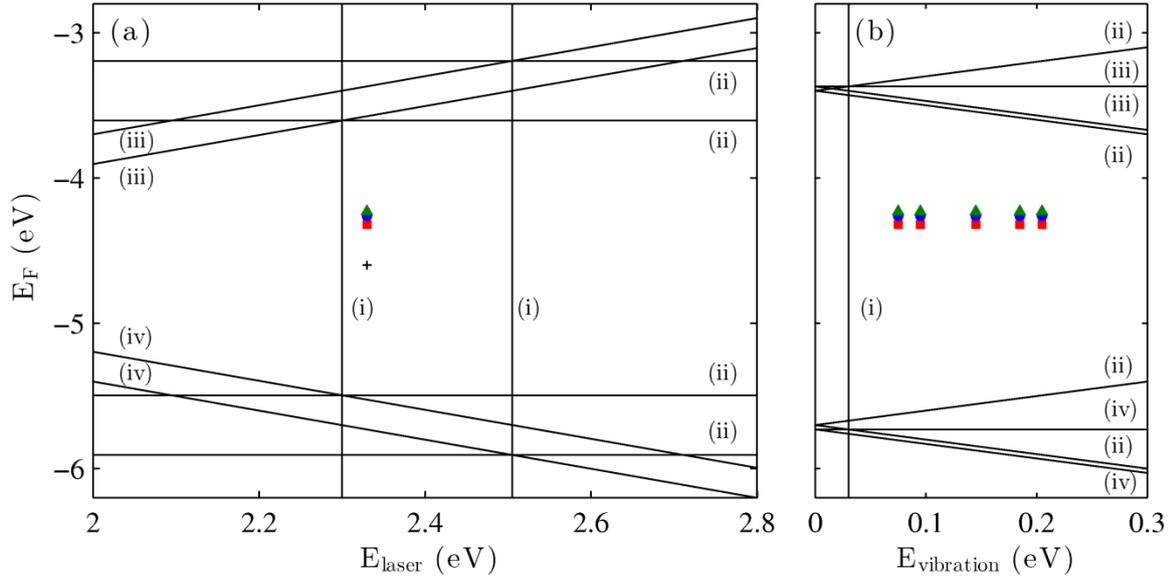

Figure 5. Graphical depiction of the enhancement conditions from Eq. 1 labeled with the number of corresponding equation. Panel (a) represents the data in the phase space of the energy of the laser and the Fermi energy of graphene. In (b), the representation in the phase space of the energy of the R6G vibrations and the Fermi energy of graphene is displayed. The '+' sign represents the predicted Fermi energy of ideal undoped graphene, and the red square, blue circle and green triangle stand for fluorinated, pristine and 4-nitrophenyl functionalized graphene, respectively.

Figure 5(a) shows, in the phase space of the laser energy and the Fermi energy an example of how the conditions of Equation 1 are fulfilled for the R6G band at 1650 cm$^{-1}$ (with vibrational energy of 205 meV) in differently functionalized graphene samples. In Figure 5(b), the points correspond to different Raman bands of R6G (with different vibrational energies) for differently functionalized graphene samples at the constant laser excitation energy of 2.33 eV. We note that the lines used in Figure 5 are simplifications. The theoretical calculations do not lead to ideal lines, but they are blurred with some width profiles [10]. Therefore, the conditions do not have to be fulfilled perfectly to achieve the enhancement of the signal. Nevertheless, the better matched the conditions are, the stronger is the obtained enhancement. The first important condition is fulfilled if optimum laser excitation energy is chosen (resonant Raman scattering, condition (*i*) from Equation 1). From both panels in Figure 5, it can be observed that with increasing doping, the system comes closer to the lines defined by conditions (*ii*) and (*iii*) of Equation 1. This result is in agreement with the relative enhancement obtained for differently functionalized graphene samples. The highest enhancement is found for the 4-nitrophenyl functionalized graphene, which achieves the highest doping, while the fluorinated graphene, which is the least doped, shows the lowest enhancement of the R6G Raman signal. From panel (b) of Figure 5, it is expected that with the increasing vibrational energy, the enhancement should be higher because of a better match with conditions (*ii*) and (*iii*) of Equation 1. This is indeed observed (Figure 4), since for higher vibrational energies, the relative enhancement was stronger for both the pristine and 4-nitrophenyl functionalized samples compared to fluorinated graphene. It is important to note that the dependences in Figure 4(b,c) are related to the fluorinated graphene, which had the lowest doping and also provides the weakest enhancement of the R6G signal. However, as the fluorinated graphene provides the weakest enhancement, i.e., the enhancement is the least affected by conditions (*ii*) and (*iii*) of Equation 1, the fluorinated provides a suitable reference for the qualitative description of the more doped pristine and 4-nitrophenyl functionalized graphene samples.

**5. Probing more excitation energies**

To further validate the theoretical concept presented in [10], we measured the GERS effect from the R6G molecules while varying the excitation energy of the laser. Apart from the laser energy of 2.33 eV that was discussed above, we tested laser energies of 1.96, 2.18, 2.41, 2.54 and 2.71 eV. The GERS spectra of the pristine graphene samples for individual excitation energies are shown in Figure 6. For the laser energies that are too far from the resonance condition of the R6G molecules, we did not obtain any measurable GERS signal. This was the case for laser energies of 2.71 eV (too high) and 1.96 eV (too low). In contrast, for the laser excitation energies of 2.18, 2.33 and 2.41 eV, the GERS of the R6G molecules can be easily observed. The quantitative comparison of the enhancement at different laser excitations is difficult. However, this result demonstrates the breadth of the resonance window. The spectrum measured with the laser excitation energy of 2.54 eV provided even more insight into the mechanism behind GERS. Here, the Raman bands with high vibrational energies show enhancement, while the Raman bands with low vibrational energies do not. Namely, the Raman band at 1650 cm$^{-1}$ is clearly visible, while there are no bands at 610 or 775 cm$^{-1}$. Nevertheless, considering the conditions in Equation 1, it is possible to rationalize the latter observation.

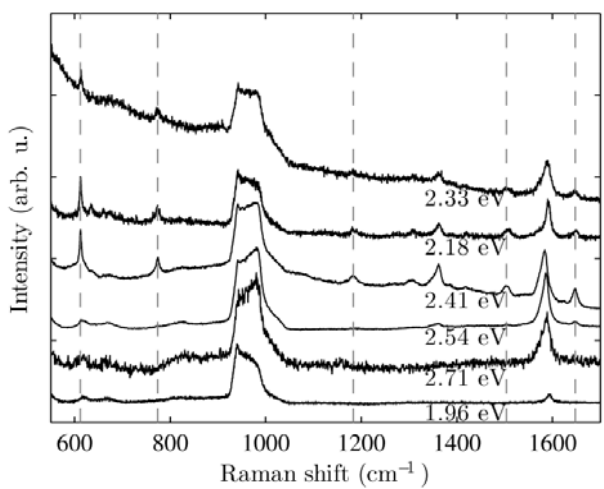

Figure 6. Raman spectra of the R6G molecules on pristine graphene measured with different laser energies. The dashed gray lines show the positions of the investigated Raman bands of the R6G molecules.

In Figure 7, we show how the resonance conditions depend on the laser excitation energy in the phase space of the laser energy and Fermi energy of graphene. Figure 7(a) is for the vibrational energy of 0.205 eV (1650 cm$^{-1}$), and Figure 7(b) is for 0.075 eV (610 cm$^{-1}$). The plots in Figure 7 rationalize the changes in the enhancement of the signal of the R6G molecules using excitation with the laser energy of 2.54 eV. While the point representing the vibrational energy of 0.205 eV is close to the lines representing conditions (*i*) and (*ii*) of Equation 1, the point representing the vibrational energy of 0.075 eV is moved away from the lines representing these conditions, and the enhancement of the vibrational energy of 0.075 eV is consequently expected to be weak. This is in agreement with the measured GERS data. The enhancement of the R6G Raman signal measured using other laser excitation energies can be explained by matching the resonant condition of the R6G molecule. Consequently, it follows from our data that a prerequisite for the observation of the enhancement is the fulfillment of the resonant condition (*i*). In other words, when the laser excitation energy is far from the difference between the HOMO and LUMO energies, no GERS was observed. For example, even though the laser energy of 1.96 eV matches condition (*iii*) of Equation 1, we observed no GERS signal because the laser energy is far from the resonance of the R6G molecules. Conditions (ii-iv) also influence the enhancement of the signal, but they do not necessarily need to be fulfilled to observe the GERS of R6G.

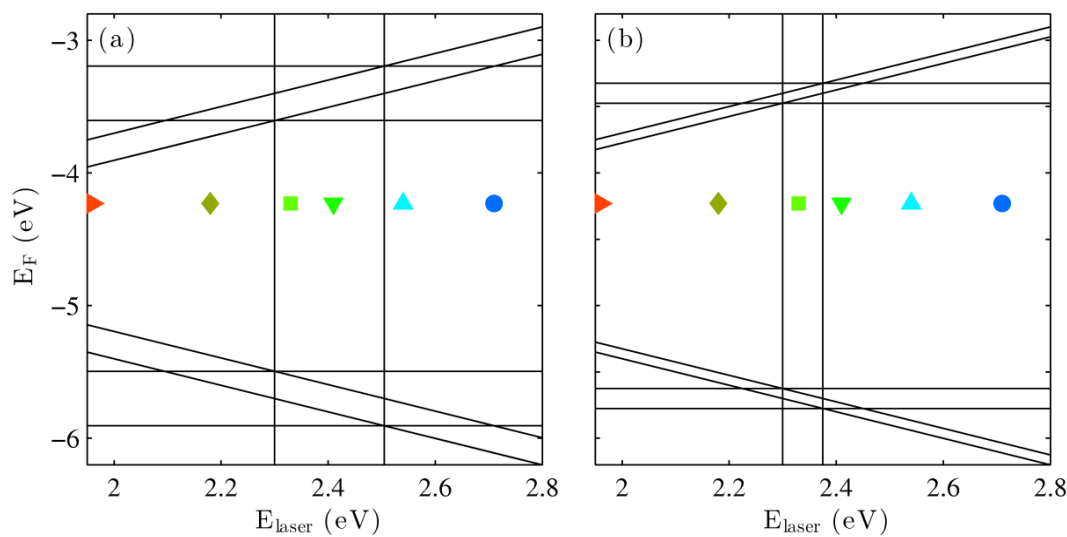

Figure 7. Graphical depiction of the enhancement conditions from Eq. 1 including the points corresponding to the used laser energies of 1.96, 2.18, 2.33, 2.41, 2.54, and 2.71 eV. The graphs were calculated for the vibrational energies of 0.205 eV (1650 cm$^{-1}$, (a)) and 0.075 eV (610 cm$^{-1}$, (b)).

## 6. Conclusions

The GERS performance of pristine, fluorinated and 4-nitrophenyl functionalized graphene substrates was studied using Rhodamine 6G probe molecules. The Raman bands of the R6G molecules exhibited significant intensities on all the studied graphene substrates, while on a control substrate (bare Si/SiO$_2$), we did not observe any traces of Raman scattering from the R6G molecules. We considered both the impact of the Fermi energy of graphene and the vibrational energy of the molecules in order to explain the GERS effect. It was found that with the increasing doping level of the graphene substrate, induced by the functional groups, the enhancement increases, which is in agreement with previous theoretical calculations. This was confirmed independently by rationalizing the enhancement by the Fermi energy of graphene and vibrational energy of the molecules. The proposed method is based on comparing the relative intensities of multiple molecular Raman modes for differently doped graphene substrates at the same time. For this reason it is possible to validate the theoretical model without knowing the absolute amount of the R6G molecules on the sample and without knowing the enhancement factors. This is highly important for molecules that exhibit photoluminescence. The highest enhancement was observed for the 4-nitrophenyl functionalized graphene sample. Furthermore, we probed the dependence of the enhancement of the Raman signal on the laser excitation energy and energy of the probed vibration which is in agreement with the theory as well.

**Acknowledgement**

The authors acknowledge support from the GACR project No.:15-01953S.